\documentclass{PoS}
\usepackage{graphicx}
\title{
 Non-locality of the nucleon-nucleon 
  potential from lattice QCD
}
\ShortTitle{
 Non-locality of the NN potential from Lattice QCD
}

\author{
\speaker{Keiko Murano}\\
        KEK, Tsukuba, Ibaraki 305-0801, Japan \\
        E-mail: \email{murano@het.ph.tsukuba.ac.jp}
}
\author{Noriyoshi Ishii\\
  Department of Physics, The University of Tokyo,
  Tokyo 113--0033, Japan
  }
\author{Sinya Aoki\\
  Graduate School of Pure and Applied Sciences, University of Tsukuba,\\
  Tsukuba, Ibaraki 305--8571, Japan, and\\
  Center for Computational Sciences, University of Tsukuba, Tsukuba, Ibaraki 305-8577, Japan 
  }
\author{Tetsuo Hatsuda\\
  Department of Physics, The University of Tokyo,
  Tokyo 113--0033, Japan, and\\
  Institute for the Physics and Mathematics of the Universe (IPMU), 
  The University of Tokyo,\\
  Chiba 277-8568, Japan
   }
  \author{
  for HAL QCD Collaboration
  \quad
  \includegraphics[
  height=0.17\textwidth, bb=0 0 202 118]{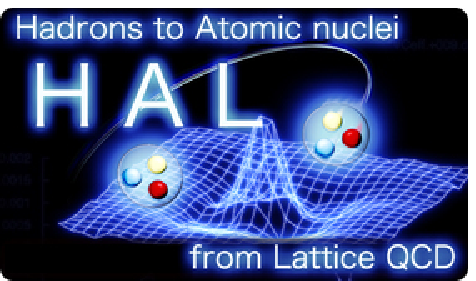} 
  } 
\abstract{The Nambu-Bethe-Salpeter (NBS) wave function
for two nucleons on the lattice has been shown to yield a non-local and 
  energy-independent nucleon-nucleon (NN) potential,  $U({\vec r},{\vec r}')$.
 In practice, the derivative expansion of $U({\vec r},{\vec r}')$ is currently
 employed
 to determine the potential at low energies.
 In this report, we  study the magnitude of
 non-locality to check the convergence of such a derivative expansion. 
 With quenched lattice QCD at $m_\pi = 530 \mbox{MeV}$,
   we  compare the NN potentials
  at the center of mass energy  $E\simeq 0$ MeV and at $E\simeq 45$ MeV.
   We also investigate
  the angular momentum dependence of the spin singlet potential,  by comparing
  the potentials in  $^1S_0$  and $^1D_2$  channels.  We find that 
  the non-locality and the angular momentum dependence 
  in the above energy range are  negligible  within statistical errors.
 }
\FullConference{The XXVIII International Symposium on Lattice Field Theory, Lattice2010\\
		June 14-19, 2010\\
		Villasimius, Italy}

\newcommand{\Eq}[1]{eq.(\ref{#1})}

\begin{document}
\section{Introduction}
 The nucleon-nucleon (NN) potential, which describes
 the interaction  between two  nucleons at low energies,
  is widely  used  as a  basic tool  to study the structures and
 reactions of atomic nuclei.
In the past few decades, several realistic NN potentials
have been proposed: These  are phenomenologically determined
so as to reproduce the NN  scattering phases of more than 4000 data points
 with $\chi^2/N_{\rm  dof}\sim 1$ \cite{Machleidt:2000ge,Wiringa:1994wb,Stoks:1994wp}.
Recently  a first-principle method
to extract the NN potential from the Nambu-Bethe-Salpeter (NBS) wave function
 in lattice QCD  has been  proposed by
 three of the present authors \cite{Ishii:2006ec,Aoki:2009ji}.
This method  can  be generalized also to derive   
 the  hyperon potentials (YN and YY) \cite{Nemura:2008sp,Inoue:2010hs}
  and the three-nucleon potentials \cite{Doi:2010yh}, which do not have
   enough experimental data.  
  The  potentials based on lattice QCD,  once obtained in good accuracy,
  would become first-principle inputs to the studies of
  ordinary nuclei, hyper nuclei, and the interior of the neutron stars
  by using advanced techniques in nuclear many-body problems.   

 The potential obtained from the NBS wave function in lattice QCD simulations  is non-local and energy-independent,  $U(\vec{r},\vec{r}')$.
 To determine the potential at low energies, we have proposed
 to make a  derivative expansion  of $U(\vec{r},\vec{r}')$
  and calculate the local coefficient functions successively \cite{Aoki:2009ji}.
 The aim of this report is  to examine the convergence of such a derivative
 expansion.
In sec.\ref{V-intro},  we give a  brief review of our  method to construct
the  non-local potential and its  derivative  expansion.
In  sec.\ref{sec:non-locality},  we show that
the  non-locality of $U(\vec r,\vec r')$ can be checked
 by the center of mass energy ($E$) dependence and the orbital angular momentum  ($L$)
   dependence  of a ``truncated local potential", 
   $V^{\rm trun}(r;E,L)$.
In  sec.\ref{sec:results}, we  show our  numerical results  of 
 $V^{\rm trun}(r;E,L=0)$ at two different scattering energies,
  $E\simeq 0$ MeV and  45 MeV.
 In addition we examine  the  $L$  dependence  of 
 the truncated local  potential in the two spin-singlet channels,
 $^1S_0$  and  $^1D_2$.
We  find  negligible energy  and  angular momentum
dependence of $V^{\rm trun}(r;E,L)$, which 
 implies that the non-locality $U(\vec{r},\vec{r}')$ is small
  at low energies up to $E \simeq 45$ MeV.

\section{NN potential from Lattice QCD and its derivative expansion}\label{V-intro}
The equal-time NBS wave function\footnote{The NBS wave function was also utilized
  to investigate L\"uscher's
 finite volume method  for the pion-pion scattering phase shift
 in Ref.\cite{Lin:2001ek,Aoki:2005uf}.}
 is defined as
\begin{eqnarray}
\phi(\vec{r},k) \equiv
\frac1{V}\sum_{\vec x}
\langle0|n(\vec{x}+\vec{r})p(\vec{x}) | B=2;k \rangle ,
\end{eqnarray}
where  "$k$" is the ``asymptotic momentum'' related to
the  total energy as $W=2\sqrt{m^2_N+k^2}$.
 Also, the local  composite operators for the nucleon are defined as
$
  n_\beta(y)
  =
  \epsilon_{abc}
  \left(u_a(y) C\gamma_5 d_b(y)\right) d_{c\beta}(y)
$
and
$
  p_\alpha(x) =
  \epsilon_{abc}
  \left(u_a(x) C\gamma_5 d_b(x)\right)
  u_{c\alpha}(x)$,  where $a,b,c$  denote color indices,  and $C$  is the
charge conjugation matrix.
It has been discussed that $\phi(\vec  r, k)$ 
 below  the pion production threshold satisfies the   Schr\"odinger-type
     equation \cite{Ishii:2006ec,Aoki:2009ji}:
 \begin{equation}
  \left(\triangle + k^2\right) \ \phi(\vec{r};k) = 2\mu \int d^3 \ r^\prime \
   U(\vec{r},\vec{r}^\prime) \ \phi(\vec{r}^\prime,k), \label{eq:SE}
\end{equation}
which defines the  non-local potential and $k$-independent potential
$U(\vec{r},\vec{r}^\prime)$ with
 $\mu\equiv m_{N}/2$ being the reduced mass.
 By using the Nishijima-Zimmermann-Haag
reduction formula for composite operators,
 it can be shown that, for large $|\vec r|$, 
 $\phi(\vec  r;k)$ has an asymptotic form  characterized
by the scattering phase $\delta(k)$  \cite{Aoki:2009ji}:
\begin{equation}
  \phi(\vec r; k)
  \sim
  \frac{\sin(kr + \delta(k))}{kr} + \cdots.
\end{equation}
This functional form is exactly the  same as the asymptotic form of the
scattering  wave in  the non-relativistic  quantum  mechanics. Therefore
our non-local potential $U(\vec r,\vec  r')$ is to reproduce the scattering phase 
$\delta(k)$ below the pion production threshold.
 
Although  the  non-local potential  $U(\vec  r,\vec  r')$  has nice formal
properties, its  explicit construction is not easy in  lattice QCD simulations, where
 the number of available  NBS  wave functions  for different 
 scattering energies are limited.
 Therefore, we  employ the derivative expansion as shown below and
 determine the local coefficient 
 functions successively by using low-energy NBS wave functions: 
\begin{equation}
  U(\vec{r},\vec{r}^\prime)
  = \bigg[
    \underbrace{V_{0}^I(r)
      +
      V_\sigma^I(r) \
      \left(\sigma_1\cdot\sigma_2\right) + V_T^I(r) \
      S_{12}
    }_{LO}
    +
    \underbrace{
      V_{LS}^I(r) \ \vec{L}\cdot \vec{S}
    }_{NLO}
    + \mathcal{O}(\vec{\nabla^2})
    \bigg]\delta(\vec{r}-\vec{r}^\prime).
  \label{eq:delive}
\end{equation}
Here $I=0,1$  denotes the  total isospin of  the two nucleons and
$\vec{S}\equiv(\vec{\sigma}_1+\vec{\sigma}_2)/2$  is  the  total
spin operator.  Also, the tensor operator is written as 
$S_{12} \equiv 3  (\vec{\sigma}_1 \cdot \vec{r}) (\vec{\sigma}_2 \cdot
\vec{r})  /   r^2  -  \vec{\sigma}_1   \cdot  \vec{\sigma}_2$.
 The LO terms such as 
 the central potential in the $^1S_0$ channel, and the central
and the  tensor potentials  in the  $^3S_1-^3D_1$  channel 
 have been obtained  so far at CM energy $E\sim 0$ MeV  
 (see e.g. a recent summary  \cite{Ishii:2010th, Aoki:2009ji}).

\section{Energy and angular momentum dependence as a test of non-locality}
\label{sec:non-locality}

In this section we discuss that  the  convergence  of  the  derivative
expansion of $U(\vec{r},\vec{r}')$ in the NN system can be examined 
 by studying the energy and  the angular momentum dependence  of
  the truncated local potential at finite $E$.
\footnote{In  the Ising  field theory,  it is  analytically  shown that  the
energy-dependence  is weak  at low  energy,  indicating  that the
non-locality of the potential is weak \cite{Aoki:2008yw}.}
  For simplicity, we restrict ourselves to
the spin-singlet case ($S=0$) where  
$S_{12}$ and $\vec L\cdot\vec  S$ do not contribute. 
 Then we obtain
\begin{eqnarray}
  \left(\Delta + k^2\right)
  \phi^{S=0}(\vec{r};k)
  =
  &2\mu
  &\left[
    V_{\rm C}(r)
    + \left\{ \nabla^2, V_{p^2}(r)    \right\}
    + V_{L^2}(r)    \ L^2
    + \cdots
    \right]
      \phi^{S=0}(\vec{r};k),
    \label{eq:S0}
\end{eqnarray}
where $V_{\rm C}(r)  \equiv   V_{0}(r) -3V_{\sigma}(r)$ is the LO potential in
the  derivative expansion, while
 the higher order terms  with $L^{2n}$, $\nabla^{2n}$ are referred to  
 N$^{2n}$LO terms.
 If we have independent  NBS wave functions for different 
  energies and angular momenta, we can determine 
  N$^{2n}$LO terms order by order starting from a small $n$.
 For such a procedure making sense, the magnitude of the 
  potential at the  leading order should be dominant
   at low energies and low angular momenta.
    A simplest way to check this is to  define
  a ``truncated local potential" given below  and study
 its $E$ and  $L$  dependence:
\begin{equation}
  V_{\rm  C}^{\rm trun}(r;E,L)
  \equiv
  \frac1{2\mu}
  \frac{\nabla^2 \phi^{S=0}(\vec{r};k)}
  {\phi^{S=0}(\vec{r};k)}
  +
  \frac{k^2}{2\mu}.
  \label{eq:VCLO}
\end{equation}
If  $V_{\rm  C}^{\rm trun}(r;E,L)$ has small $E$ and $L$ dependence,
 it is a good approximation of  $V_{\rm  C}(r)$. On the other hand,
 if it has  large  $E$ and/or $L$ dependence, one cannot
  neglect  $V_{p^2}(r)$, $V_{L^2}(r)$  etc, so that
  they must be determined together with 
   $V_{\rm  C}(r)$ by using the NBS wave functions
for several different values of $E$ and $L$.

\section{Numerical Simulations and results}\label{sec:results}
\subsection{Lattice QCD setup}
We  employ the  standard plaquette  gauge action  on  a $32^3\times48$
lattice  at $\beta=5.7$  for quenched  gauge  configurations.  Quark
propagators  are calculated  with the  standard Wilson  quark  action at
$\kappa=0.1665$.    This   setup   gives  the   lattice   spacing
$a^{-1}=1.44(2)$  GeV  ($a\simeq 0.137$  fm)  from  $m_\rho$  input,  the
spatial extension $L=32a \simeq 4.4$  fm, $m_\pi \simeq 0.53$ GeV and $m_N
\simeq 1.33$ GeV  \cite{Fukugita:1994ve}.  Quenched gauge configurations
are  generated  by  the  heat  bath algorithm  with  over  relaxations.
Potentials are  measured with configurations separated by  $200$ sweep, and
$4000$  configurations  are  accumulated  to obtain  results  in  this
report. These calculations are performed on Blue Gene/L at KEK.

The  NBS  wave  function   is  obtained  from  the  nucleon  four-point
correlator in the large $t$ region,
\begin{eqnarray}
  G^{(4)}(\vec{x},\vec{y},t,t_0)
  &=&
  \langle 0 |
  n(\vec{y},t)
  p(\vec{x},t)
  \bar{\mathcal{J}}_{pn}(t_0)
  |0\rangle 
  =
  \sum_n
  A_n
  \langle 0|
  n(\vec{y})
  p(\vec{x})
  |B=2;W_n\rangle
  e^{-W_n(t-t_0)}
  \nonumber \\
  &\rightarrow& 
  A_0
  \langle 0|
  n(\vec{y})
  p(\vec{x})
  |B=2;W_0\rangle
  e^{-W_0(t-t_0)},
  \label{eq:4-point} 
\end{eqnarray}
where   $W_n   =   2\sqrt{m_N^2+k_n^2}$   and  $A_n   \equiv   \langle
B=2;W_n| \bar{\mathcal{J}}_{pn}|0 \rangle$.
Here $\bar{\mathcal{J}}_{pn}(t_0)          \equiv         \overline{P}(t_0)
\overline{N}(t_0)$  is a  source operator located at  $t=t_0$
 with  $\bar{P}_\alpha  \equiv  \epsilon_{a,b,c}  \left(  \bar{U}_a  C
\gamma_5  \bar{D}_b   \right)  \bar{U}_{c\alpha}$  and  $\bar{N}_\beta
\equiv \epsilon_{a,b,c} \left(  \bar{U}_a C \gamma_5 \bar{D}_b \right)
\bar{D}_{c\beta}$.  Up and down quark operators
associated with a  source function $f(x)$ specified later  are denoted as 
$U(t) \equiv  \sum_{\vec{x}} u(t,\vec{x})f(\vec{x})$ and  $D(t) \equiv
\sum_{\vec{x}} d(t,\vec{x})f(\vec{x})$.
By  examining the  $t$ dependence  of potentials,  we confirm  that the ground
state saturation for potentials is achieved at $t-t_0\ge 9$.

\subsection{Details of the calculation}
The NN potentials at CM energy  $E\equiv k^2/(2\mu) \simeq 0$ MeV 
 with $\mu=m_N/2$ is constructed
by using  the periodic boundary  condition (PBC), which is  imposed on
the quark  fields along spatial directions.  Since spatial
momentum of  each quark is discretized as  $p_i \simeq  2\pi n_i/L$
($n_i \in \mathbb{Z})$ in the PBC, the ground state corresponds to $p_i \simeq 0$.
For the PBC, we  employ the wall source, i.e., $f(\vec r)  = 1$ with the Coulomb gauge fixing only at $t=t_0$ to enhance
the overlap  with the ground  state. 
The  wall source  provides us  with the $A_1^+$  orbital  contribution, by
which we  can only study  S-wave ($L=0$)  for  the spin singlet case with
possible contaminations from higher partial waves with $L\ge 4$.
To improve  the statistics, we  calculate the NBS wave function four times on each configuration by putting the source point at  four different time slices.

The NN  potentials at  CM  energy $E\neq  0$  is constructed  by using  the
anti-periodic boundary condition (APBC), which is imposed on the quark
fields along spatial directions.  Since the spatial momentum of
each  quark
  is discretized as $p_i  \simeq (2n +  1)\pi/L$ ($n_i \in
\mathbb{Z}$) in the APBC, the ground state  corresponds to $p_i \simeq \pm \pi/L$,
whose energy  is given by  $E \simeq 3\times (\pi/L)^2/(2\mu)  \simeq 45$
MeV for $L\simeq 4.4$ fm.
For the APBC, we employ four types of ``momentum-wall" sources, i.e., $f(\vec
r)        \equiv        \cos((x+y+z)\pi/L),       \cos((-x+y+z)\pi/L),
\cos((x-y+z)\pi/L)$, and $\cos((-x-y+z)\pi/L)$  with the Coulomb gauge fixing only at $t_0$ to enhance the overlap
with the ground state with $p_i \sim \pm \pi/L$.
In addition to the $A_1^+$  orbital contribution, our momentum-wall source
provides us with the $T_2^+$ orbital contribution, which  makes  it  possible  to  study  D-wave  ($L=2$)   for spin singlet case with  possible contaminations from higher partial waves with $L\ge 4$.

The  NBS wave  functions for  the S-wave and  the D-wave are  extracted  by
projection operators,
\begin{eqnarray}
  \phi(\vec{r};k;\Gamma) = \frac{d_\Gamma}{24}
  \sum_{i=0}^{23}\chi^{(\Gamma)}(R_i)^* \ \phi^{S=0}(R \cdot \vec{r};k),
\end{eqnarray}
where  $\chi^{(\Gamma)}$ denotes  the character  of  the representation
$\Gamma$ for  the cubic  group $O$ (See  Table.\ref{Table:chi}),
$R_i$ denotes one of the 24 elements of the cubic group, and 
$d_\Gamma$   denotes  the  dimension   of  the   representation,  i.e.,
$d_{A1}=1$ and $d_{T2}=3$.
\begin{table}[b]
\begin{center}
 \begin{tabular}{|c||ccccc|}
  \hline
  $\Gamma$ & E & $6C_4$ & $3C_2$ & $8C_3$ & $6C_2$\\
  \hline
  $A_1$ & 1 &  1 &  1 &  1 &  1 \\
  $A_2$ & 1 & -1 &  1 &  1 & -1 \\
  $E$   & 2 &  0 &  2 & -1 &  0 \\
  $T_1$ & 3 &  1 & -1 &  0 & -1 \\
  $T_2$ & 3 & -1 & -1 &  0 &  1 \\
  \hline
 \end{tabular}
 \caption{The character table of the cubic group $O$.  The notations of
 classes     ($E,6C_4,\cdots$)    and     irreducible    representation
 ($A_1,A_2,\cdots$) follow Mulliken Symbols.}
 \label{Table:chi}
 \end{center}
\end{table}
The calculation performed with $\Gamma=A_1$ in  the PBC provides us with the NBS  wave function for $^1S_0$  at $E \simeq 0$ MeV, while the calculation with $\Gamma=A_1$ and $T_2$ in the APBC provides us with NBS wave functions for $^1S_0$  and $^1D_2$ at $E\simeq 45$ MeV, respectively.
These  NBS wave  functions  are inserted  into  \Eq{eq:VCLO} to  obtain
$V_{\rm C}^{\rm trun}(r;E,L)$,  where  $k^2/2\mu$ in  this report is estimated from the 
 non-interacting nucleons on the lattice,     i.e., 
     $k^2/2\mu=0$     MeV    for   the PBC,     and
$k^2/2\mu=3(\pi/L)^2/2\mu=45$ MeV for the APBC. 

\subsection{Results}
\begin{figure}
\begin{center}
 \scalebox{0.85}{\includegraphics{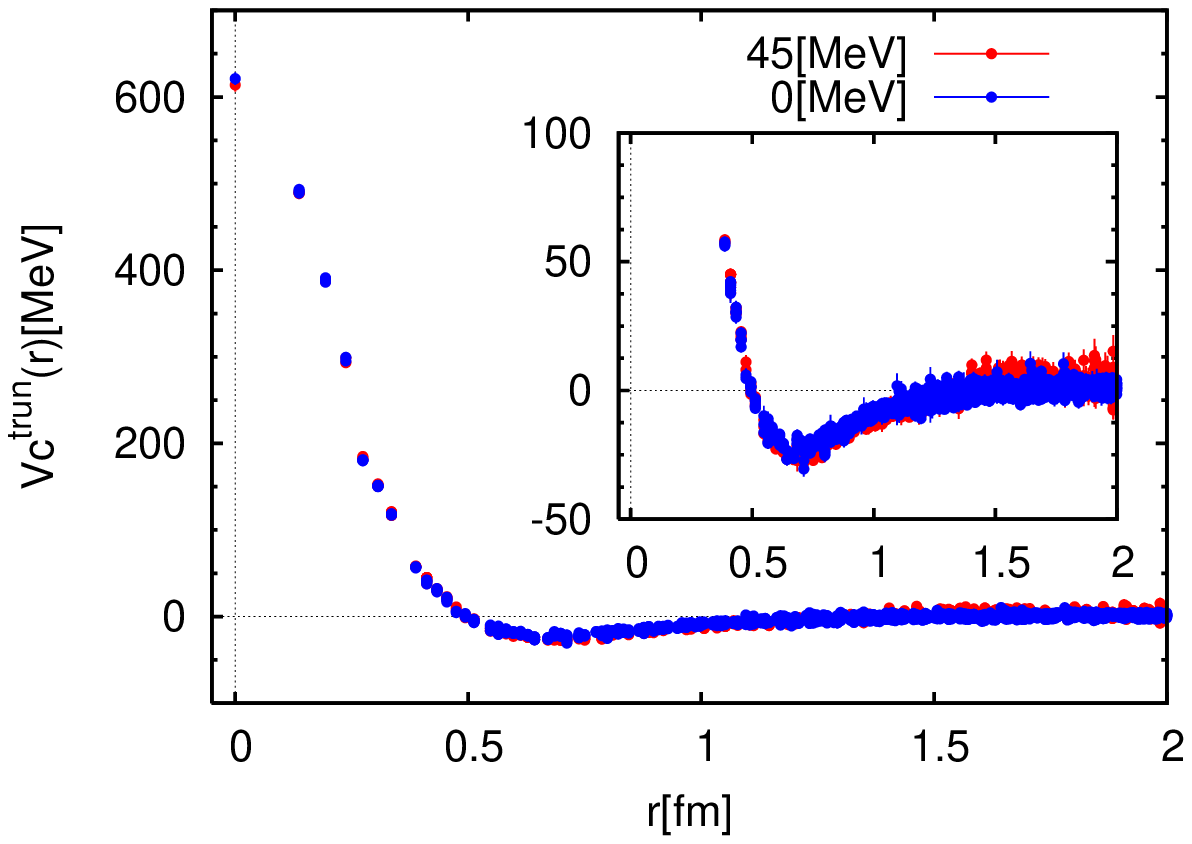}} \\
 \scalebox{0.85}{\includegraphics{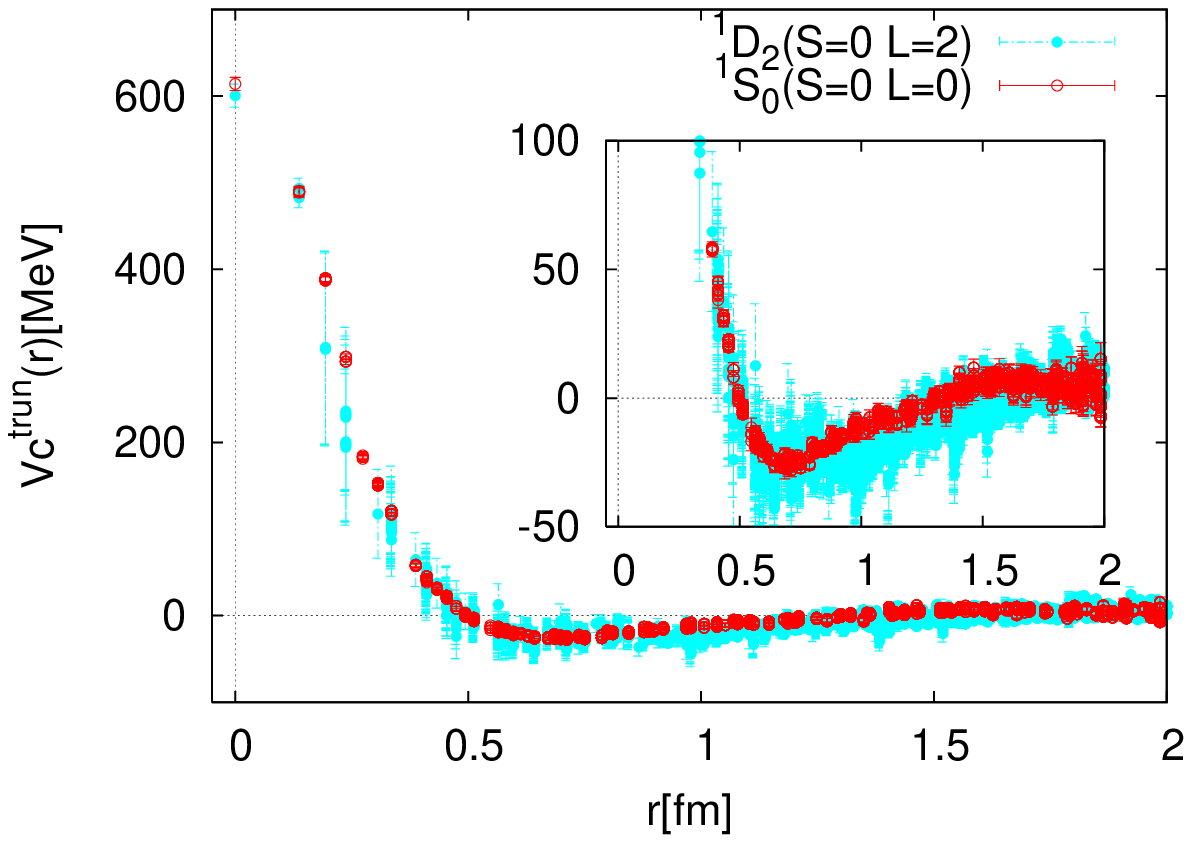}}
\end{center}
 \caption{Top: A comparison of 
 the truncated local potential $V_{\rm C}^{\rm trun}(r;E,L=0)$ in the $^1S_0$  channel
   with  PBC (blue) and  that with APBC (red) at $t-t_0=9$ and $r \le 2.0$
  fm.  Bottom: A comparison  of $V_{\rm C}^{\rm trun}(r;E,L)$
  with the APBC  for the $^1S_0$ channel  (red) and that for 
  the $^1D_2$ channel (cyan).  }
  \label{fig:APBC_vs_PBC}
\end{figure}

In  the top panel  of  Fig.  \ref{fig:APBC_vs_PBC},  we  compare 
 $V_{\rm C}^{\rm trun}(r;E,L=0)$ in the 
$^1S_0$ channel  obtained
at $E\simeq 45$  MeV (red) with that at $E\simeq  0$ MeV (blue).  All
data  are  taken at  $t-t_0=9$,  where  the ground  state saturation  is
numerically confirmed as mentioned before.   We observe  that  the  agreement of 
the  potentials  between  the two energies is quite good within  statistical errors.
The comparison of $V_{\rm C}^{\rm trun}(r;E,L)$ in the  $^1S_0 (L=0)$ channel
 (red) with that in the  $^1D_2 (L=2)$
 channel (cyan) obtained at $E \simeq 45$ MeV is  shown in the bottom panel
 of Fig.  \ref{fig:APBC_vs_PBC}.  We observe that the angular momentum
dependence of $V_{\rm C}^{\rm trun}(r;E,L)$  is weak, though the statistical errors are rather large in the $^1D_2$ channel.

By  these comparisons, we  find  that the energy  and the angular  momentum
dependence of  $V_{\rm C}^{\rm trun}(r;E,L)$  is very weak within statistical errors.
We therefore conclude that  contributions  from  higher  order  terms  in  the  derivative
expansion are small and that  the LO local potential in the derivative expansion is the
 good    approximation     for     the    non-local     potential
 $U(\vec{r},\vec{r}')$ at least up to the energy $E \simeq 45$ MeV and
 angular momentum $L\le 2$.

\section{Summary and discussion}\label{sec:conclusion}
We  have  studied the validity  of  the derivative  expansion  for  the
non-local  NN  potential  $U(\vec{r},\vec{r}')$  in  quenched  QCD  at
$m_\pi\simeq 530$ MeV.
The  convergence is tested by  the energy  and the  angular momentum
dependence  of the truncated local potential  $V_{\rm C}^{\rm trun}(r;E,L)$: 
To examine  the energy dependence,  we have generated  $V_{\rm C}^{\rm trun}(r;E,L=0)$ 
 in the $^1S_0$ channel at two different energies,  $E\simeq 0$ MeV and $E\simeq
45$ MeV.
We have found  that the energy dependence of the truncated  potential is
very weak.
We have also generated $V_{\rm C}^{\rm trun}(r;E,L=2)$ 
 for $^1D_2$ at $E\simeq 45$ MeV in
the same setup.  From the  comparison with  $V_{\rm C}^{\rm trun}(r;E,L=0)$  at
$E\simeq 45$ MeV, we have observed that the angular momentum dependence of
the effective potential is also small.
These results  indicate that the LO  potentials 
  constructed from NBS wave functions obtained
at $E\simeq 0$ MeV  can be safely used for $0\le E \le 45$ MeV and the angular
 momentum $L \le 2$.

An extension of the  current analysis to NN potentials in the spin-triplet ($S=1$)
 channel is ongoing.  It is an important future task
   to perform the same analysis  for 
 other baryon-baryon potentials and for the potentials in full QCD
 with lighter  pion masses.  Eventually, we need to
 extract NLO terms such as the spin-orbit potential
 from the NBS wave functions with finite  angular momentum on the lattice. 
  
\section*{Acknowledgement}
We  are grateful  for authors  and maintainers  of  CPS++\cite{cps}, a
modified version  of which is used  for simulation done  in this work.
Numerical calculations in this study have been performed under  the  Large  Scale  Simulation  Program No.0923(FY2009)  of  High  Energy  Accelerator  Research  Organization
(KEK).  This  work is  supported in part by Grant-in-Aid  of the  Ministry of
Education, Science and Technology, Sports and Culture (Nos.
20340047,  
20105001,  
20105003,  
22540268,  
19540261,  
).

\end{document}